\documentclass[aps,prb,twocolumn,eqsecnum,amsmath,amssymb,floatfix,superscriptaddress]{revtex4}
\usepackage{graphicx}
\usepackage{subfigure}
\usepackage{psfrag}
\usepackage{amsmath}
\usepackage{dcolumn}
\usepackage{bm}
\usepackage{hyperref}
\usepackage{tabularx}
\usepackage{slashbox}

\newcommand{\vp}{\varphi}

\newcommand{\rr}{\mathbf{r}}
\newcommand{\kl}{\kappa_L}
\newcommand{\kt}{\kappa_T}
\newcommand{\verts}[1]{\left\vert #1 \right\vert}
\newcommand{\ud}{\mbox{d}}
\newcommand{\degree}{\ensuremath{^\circ}}

\def\be{\begin{equation}}
\def\ee{\end{equation}}
\def\ba{\begin{eqnarray}}
\def\ea{\end{eqnarray}}

\def\LSCO{La$_{2-x}$Sr$_x$CuO$_4$}
\def\LCO{La$_2$CuO$_4$}

\def\LBCO{La$_{2-x}$Ba$_x$CuO$_4$}

\def\C60{A$_x$C$_{60}$}
\def\LNSCO{La$_{1.6-x}$Nd$_{0.4}$Sr$_x$CuO$_{4}$}

\def\HgCu3{HgCa$_2$Cu$_3$O$_{8+y}$}
\def\HgCu4{HgBa$_2$Ca$_3$Cu$_4$O$_{10+y}$}
\def\TlCu3{Tl$_2$Ba$_2$Ca$_2$Cu$_3$O$_{10+y}$}
\def\TlCu4{Tl$_2$Ba$_2$Ca$_3$Cu$_4$O$_{12+y}$}

\def\BiCu3{Bi$_2$Sr$_2$Ca$_{2}$Cu$_3$O$_y$}

\begin{document}

\title{Distinguishing Patterns of Charge Order:  Stripes or Checkerboards}

\author{John A. Robertson}
\affiliation{Department of Physics, Stanford University, Stanford, CA 94305-4045, USA}

\author{ Steven A. Kivelson}
\affiliation{Department of Physics, Stanford University, Stanford, CA 94305-4045, USA}

\author{Eduardo Fradkin}
\affiliation{Department of Physics, University of Illinois at Urbana-Champaign,
1110 West Green Street Urbana, IL 61801-3080, USA}

\author{ Alan C. Fang}
\affiliation{Department of Applied Physics, Stanford University, Stanford, CA 94305-4045, USA}

\author{Aharon Kapitulnik}
\affiliation{Department of Physics, Stanford University, Stanford, CA 94305-4045, USA}
\affiliation{Department of Applied Physics, Stanford University, Stanford, CA 94305-4045, USA}

\vspace{0.5cm}

\date{\today}

\begin{abstract}

In two dimensions, quenched disorder always rounds transitions involving the
breaking of spatial symmetries so, in practice, it can often be difficult to
infer what form the symmetry breaking would take in the ``ideal,'' zero disorder
limit.  We discuss methods of data analysis which can be useful for making such
inferences, and apply them to the problem of determining whether the preferred
order in the cuprates is ``stripes'' or ``checkerboards.''  In many cases we
show that the experiments clearly indicate stripe order, while in others (where
the observed correlation length is short), the answer is presently uncertain.
\end{abstract}

\maketitle

\section{Introduction}

Charge ordered states are common in strongly correlated materials, including
especially the cuprate high temperature superconductors.  Identifying where
such phases occur in the phase diagram, and where they occur as significant
fluctuating orders is a critical step in understanding what role they play
in the physics, more generally.  Since ``charge ordered''
refers to states which spontaneously break the spatial symmetries of the
host crystal, identifying them would seem to be straightforward.
However, two real-world issues make this less simple than it would
seem.  In the first place, quenched disorder (alas, an unavoidable presence in real
materials), in all but a very few special circumstances, rounds the transition and
spoils any sharp distinction between the symmetric and broken symmetry states.
Moreover, the charge modulations involved tend to be rather small in magnitude,
and so difficult to detect {\it directly} in the obvious experiments, such as
X-ray scattering. 
 
In a previous paper,~\cite{kivelson1} three of us addressed at some length the
issue of how the presence or absence of charge order or incipient charge order can
best be established in experiment.  
In the present paper we focus on a related issue:  in a system in which
charge order is believed to exist, how can the precise character of the charge
order best be established?  This is particularly
timely given the spectacular developments in scanning tunneling microscopy
(STM) which produces extremely evocative atomic scale ``pictures'' of the 
local electronic structure -- the question is how to extract unambiguous
conclusions from the 
cornucopia~\cite{hoffman1,hoffman2,howald1,howald2,mcelroy1,mcelroy2,momono1,hanaguri1}
of data.  
We take as a representative example, the issue of whether the charge order
that is widely observed in the cuprates is ``stripes'' (which in addition
to breaking the translation symmetry, breaks various mirror and discrete
rotation symmetries of the crystal) or ``checkerboards'' (an order which
preserves the point-group symmetries of the crystal). 
To address this issue, we generate simulated STM data and then test
the utility of various measures we have developed for discriminating 
different types of order by applying them to this simulated data.
Where the correlation length for the charge order is long, definitive
conclusions can be drawn relatively simply - consequently, it is
possible to conclude that the preferred charge order in the 214 
({\LCO}) family of materials is stripes and not checkerboards.\cite{tranquada4}   
However, where the correlation length is short (disorder effects
are strong), it turns out (unsurprisingly) to be very difficult to
develop any fool-proof way to tell whether the observed short-range
order comes from pinned stripes or pinned checkerboards -- for example,
the image in Fig.~\ref{fig1} (right panel) corresponds to disorder-pinned stripes,
despite the fact that, to the eye, the pattern is more suggestive of
checkerboard order (with the latter seen in Fig.~\ref{fig2} (right panel)).

In Section~\ref{general}, we give precise meaning in terms of broken
symmetries to various colloquially used descriptive terms such as
``stripes,'' ``checkerboards,'' ``commensurate,'' ``incommensurate,''
``diagonal,'' ``vertical,''  ``bond-centered,'' and ``site-centered.''
In Section~\ref{landau} we write an explicit Landau-Ginzburg (LG) free energy
functional for stripe and checkerboard orders, including the
interactions between the charge order and impurities.  In Section~\ref{simulated}
we generate simulated STM data by minimizing the
LG free energy in the presence of disorder.  (See Figs.~\ref{fig1} and \ref{fig2}.)
The idea is to develop strategies for solving the inverse problem:
Given the simulated data, how do we determine whether the 
``ideal'' system, in the absence of disorder, would be stripe or
checkerboard ordered, and indeed, whether it would be ordered at
all or merely in a fluctuating phase with a large CDW susceptibility
reflecting the proximity of an ordered state.   In Section~\ref{simulated},
we define several quantitative indicators of orientational order that
are useful in this regard, but unless the correlation length is well
in excess of the CDW period, no strategy we have found allows confident
conclusions. In Section~\ref{ortho}, we show that the response of the
CDW order parameter to various small symmetry breaking fields, such
as a small orthorhombic distortion of the host crystal, can be used
to distinguish different forms of charge order.  In Section~\ref{exp}, 
we apply our quantitative indicators to a sample of STM data in 
Bi$_2$Sr$_2$CaCu$_2$O$_{8+\delta}$ and discuss the results. In
Section~\ref{conclude} we conclude with a few general observations.

\section{General Considerations}
\label{general}
 
{\it Stripes} are a form of unidirectional charge order (see Fig.~\ref{fig1} (left panel),
characterized by modulations of the charge density at a single ordering
vector, $\bf Q$, and its harmonics, ${\bf Q}_n=n \bf Q$ with $n=$ an
integer.  In  a crystal, we can distinguish different stripe states not
only by the magnitude of $\bf Q$, but also by whether the order is commensurate
(when $|{\bf Q}|a = 2\pi (m/n)$ where $a$ is a lattice constant and $n$
is the order of the commensurability) or incommensurate with the underlying
crystal, and on the basis of whether $\bf Q$ lies along a symmetry axis or not.
In the cuprates, stripes that lie along or nearly along the Cu-O bond direction
are called ``vertical'' and those at roughly $45\degree$ to this axis are called
``diagonal.''  In the case of commensurate order, stripes
can also be classified by differing patterns of point-group symmetry breaking
- for instance, the precise meaning of the often made distinction between
so-called  ``bond-centered'' and ``site-centered'' stripes is that they each
leave different reflection planes of the underlying crystal unbroken. 
Furthermore, it has been argued that bond and site-centered stripes may be found
in the same material,~\cite{li1} and even may coexist at the same temperature.\cite{wochner1}
The distinction between bond and site-centered does not exist for incommensurate 
stripes.  If the stripes are commensurate, then $\bf Q$ must lie
along a symmetry direction, while if the CDW is incommensurate, it sometimes will not.

{\it Checkerboards} are a form of charge order (see Fig.~\ref{fig2} (left panel)
that is characterized by bi-directional charge density modulations, with a pair of
ordering vectors, $\mathbf{Q}_1$ and $\mathbf{Q}_2$ (where typically 
$\verts{\mathbf{Q}_1} = \verts{\mathbf{Q}_2}$.) Checkerboard order
generally preserves the point group symmetry of the underlying crystal if
both ordering vectors lie along the crystal axes.  In the case in which they
do not, the order is rhombohedral checkerboard and the point group symmetry
is not preserved.  As with stripe order, the wave vectors can be incommensurate
or commensurate, and in the latter case $\mathbf{Q}_j\,a = 2\pi \left(m/n,m^\prime/n^\prime\right)$.
Commensurate order, as with stripes, can be site-centered or bond-centered.

\section{Landau-Ginzburg Effective Hamiltonian}
\label{landau}

To begin with, we will consider an idealized two dimensional model in which
we ignore the coupling between layers   and take the underlying crystal to
have the symmetries of a square lattice.  We further assume that in the
possible ordered states, the CDW ordering vector lies along one of a pair
of the orthogonal symmetry directions, which we will call ``$x$'' and ``$y$''.
We can thus describe the density variations in terms of two complex scalar
order parameters, 
\begin{equation}
\rho(\rr) = \bar \rho + [\vp_1(\rr)e^{iQ_x x} + \vp_2(\rr)e^{iQ_y y} + {\rm c.c.}]
\label{rho}
\end{equation}
(For simplicity, we will take $Q_x = Q_y$ throughout.)
Note, the ``density, '' in this case, can be taken to be any scalar
quantity, for instance the local density of states, and need not mean,
exclusively, the charge density.  

To quartic order in these fields and lowest order in derivatives,  and
assuming that commensurability effects can be neglected, the most general
Landau-Ginzburg effective Hamiltonian density consistent with symmetry
has been written down by several authors:~\cite{mcmillan1,nakanishi1,sachdev1,zhang1}

\begin{eqnarray}
H_{eff}=\frac {\kl} 2 \left[|\partial_x \vp_1|^2 + |\partial_y\vp_2|^2\right]  
+ \frac {\kt} 2 \left[|\partial_y \vp_1|^2 + |\partial_x \vp_2|^2\right] \nonumber \\
+ \frac \alpha 2\left[|\vp_1|^2+|\vp_2|^2\right] +\frac{u}{4} \left[|\vp_1|^2+|\vp_2|^2\right]^2+\gamma|\vp_1|^2|\vp_2|^2 \nonumber \\ 
\phantom{.}
\end{eqnarray}

The sign of $\alpha$ determines whether one is in the broken symmetry phase ($\alpha<0$)
or the symmetric phase ($\alpha>0$), and in the broken symmetry phase, $\gamma$ determines
whether the preferred order is stripes ($\gamma >0$) or checkerboards ($\gamma <0$).
Note that for stability, it is necessary that $\gamma >-u$ and $u>0$;  if these
conditions are violated, one needs to include higher order terms in $H_{eff}$.
Without loss of generality, we can rescale distance so that $\kl=1$ and
the order parameter magnitude such that $u=1$.  For simplicity, in the
present paper, we will also set $\kl=\kt$, although the more general situations
can be treated without difficulty. The phases of this model in the absence of disorder are summarized in Table \ref{chart}.

\begin{table}
\begin{center}
\noindent
\begin{tabular}{|l||*{5}{c|}}\hline
\backslashbox[0.05\textwidth]{$\gamma$}{$\alpha$} &   \makebox[3em]{$\alpha>0$}  &   \makebox[3em]{$\alpha<0$}\\ \hline \hline \vspace{-0.01\textheight}
            &                         &                   \\  
            &  \hspace{0.02\textwidth} Symmetric\hspace{0.02\textwidth}               &  \hspace{0.01\textwidth} Broken Symmetry \hspace{0.01\textwidth}  \\ \vspace{-0.02\textheight}
            &                         &                   \\  \vspace{-0.01\textheight} 
\raisebox{0.01\textheight}{\hspace{0.02\textwidth}$\gamma >0$\hspace{0.01\textwidth}} &  $\genfrac{(}{)}{0pt}{}{{\rm Fluctuating}}{{\rm Stripes}}$  &  (Stripes) \\  \vspace{-0.0\textheight}
            &                         &                   \\ \hline \vspace{-0.01\textheight}
            &                         &                   \\ 
            &  Symmetric              &  Broken Symmetry  \\ \vspace{-0.02\textheight}
            &                         &                   \\ 
\raisebox{0.00\textheight}{\hspace{0.02\textwidth}$\gamma <0$\hspace{0.01\textwidth}} &  \raisebox{-0.01\textheight}{$\genfrac{(}{)}{0pt}{}{{\rm Fluctuating}}{{\rm Checkerboard}}$}  &  \raisebox{-0.01\textheight}{(Checkerboard)} \\ 
           &                         &                   \\ \hline 
\end{tabular}
\end{center}
\caption{\label{chart}Phases of the Landau-Ginzburg model, in the absence of disorder.}
\end{table}

Imperfections of the host crystal enter the problem as a quenched potential, $U(\rr)$:
\begin{equation}
H_{dis}= U(\rr) \rho(\rr)
\end{equation}
To be explicit, we will take a model of the disorder potential in which there
is a concentration of impurities per unit area, $\delta/a^2$ where $a$ is
the ``range'' of the impurity potential and $U_0$ is the impurity strength,
so $U(\rr) = \sum_i U_0 \Theta[a^2-(\rr-\rr_i)^2]$, where the sum is over
the (randomly distributed) impurity sites, $\rr_i$ and $\Theta$ is the Heaviside
function. We have arbitrarily taken $a$ to be 1/4 the period, 
$\lambda$, of the CDW, i.e. $\lambda \equiv 2 \pi/|{\bf Q}|$ and $a=\lambda/4$.
(This choice is motivated by the fact that, in many cases, the observed charge
order has a period $\lambda \approx 4$ lattice constants.)

\section{Analysis of the simulated data}
\label{simulated}

In this section we will show how these ideas can be used to interpret STM images in terms of local stripe order. In Ref.[\onlinecite{kivelson1}] it was shown that local spectral properties of the electron Green function of a correlated electron system, integrated over an energy range over a window in the physically relevant low energy regime, can be used as a measure of the local order. This is so even in cases in which the system is in a phase without long range order but close enough to a quantum phase transition (``fluctuating order'') that local defects can induce local patches of static order. From this point of view any experimentally accessible probe with the correct symmetry can be used to construct an image of the local order state. In applying the following method to real
experimental data, one must take as a working assumption that the image obtained
is representative of some underlying order, be it long-ranged or incipient.  This
analysis, of course, would not make sense if the data is not, at least in substantial
part, dominated by the correlations implied by the existence of an order parameter.

We generate simulated data as follows:  For a given randomly chosen configuration
of impurity sites, we minimize $H_{eff} +H_{dis}$ with respect to $\vp$.  This is
done numerically using Newton's method.  The order parameter texture is
then used to compute the resulting density map according to Eq.~\ref{rho}.  This
we then treat as if it were the result of a local imaging experiment, such as an
STM experiment.  

\begin{figure*}
 \includegraphics*[height=0.33\textwidth,
 width=0.9\textwidth]{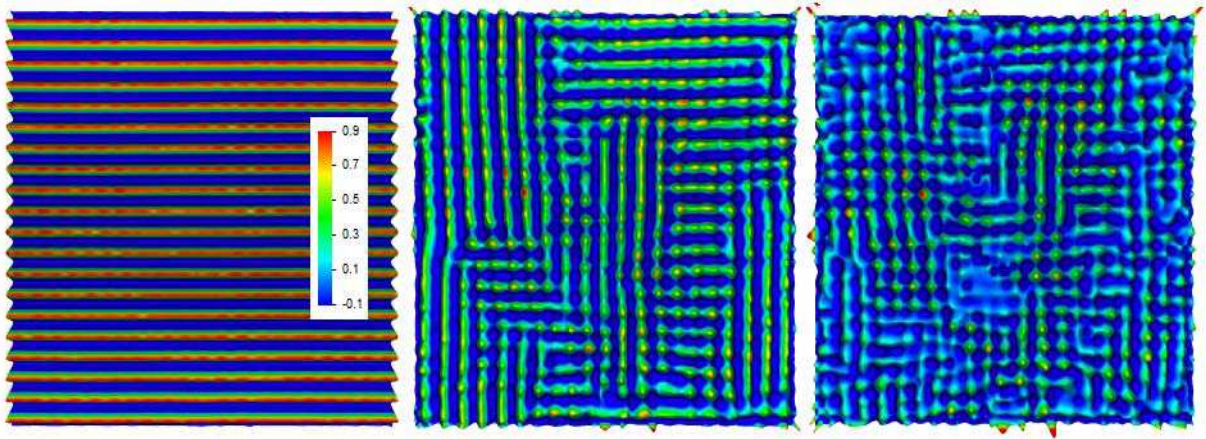}
\caption{\label{fig1}
(color online) left panel:  Highly stripe-ordered system, with weak impurities, $U_0 = 0.1$, $\delta = 0.1$.  
Here $\gamma = 1$, $\alpha = -0.05$.  [Scale is arbitrary.]
center panel:  Otherwise identical to the first system
(including the spatial distribution and concentration of impurities), but the
strength each impurities has increased to $U_0 = 0.75$. 
right panel:  Identical to the left panel, except $\alpha=+0.05$.  Much of the underlying
charge pattern remains, even to positive $\alpha$, where in the absence of
impurities, the system would be homogeneous.  All graphs are approximately 
20 CDW wavelengths in width.}  
\end{figure*}

\begin{figure*}
\includegraphics*[height=0.33\textwidth,
width=0.9\textwidth]{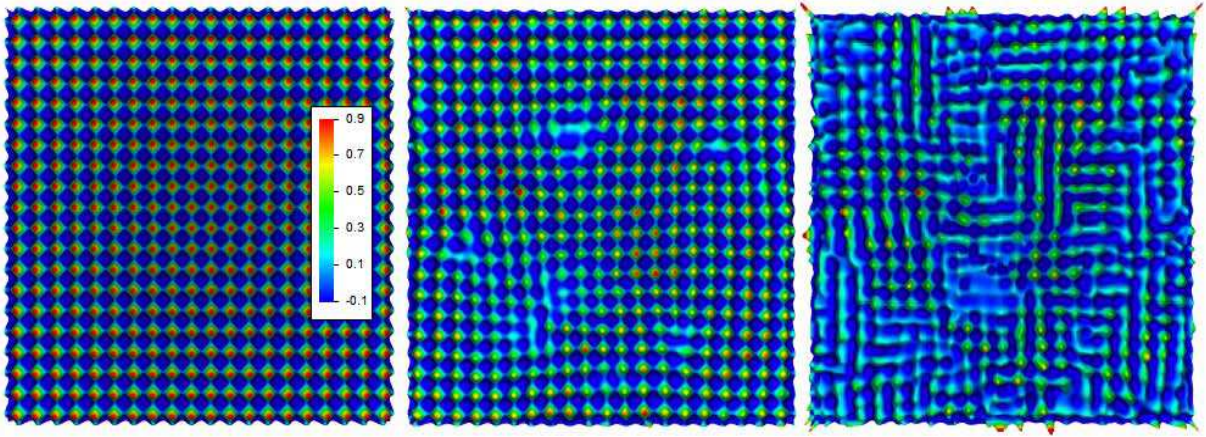}
\caption{\label{fig2} 
(color online) The parameters entering the effective Hamiltonian and the impurity realizations
are identical here to the panels of Fig.~\ref{fig1}, with the exception
of the symmetry breaking term, $\gamma$, which is now $-0.95$.
(In the center panel, because the checkerboard state is more
stable than the analogous stripe state, we have taken $U_0 = 1.5$.)
Unlike the stripe ordered system, the checkerboard system does not break into domains,
but rather develops pair wise dislocations.  In \ref{fig2} (center panel), three pairs of such dislocations are visible.
Note the similarity between the right panel of each set of Fig.\ref{fig1} and Fig.\ref{fig2}; the sign
of $\gamma$ has little effect for $\alpha>0$. 
}\end{figure*}

Even weak disorder has a profound effect on the results.  For $\alpha<0$,  
collective pinning causes the broken symmetry state to break into domains with
a characteristic size which diverges exponentially as $U_0 \to 0$ 
(In three dimensions, the ordered state survives as long as the disorder is
less than a critical value.)  Examples of this are shown in Fig.~\ref{fig1} and Fig.~\ref{fig2},
where data with a given
configuration of impurities with concentration $\delta=0.1$ are shown for
various strengths of the potential, $U_0$.  For a checkerboard phase ($\gamma < 0$),
the domain structure is rather subtle, involving shifts of the phase of
the density wave as a function of position as can be seen in Figs.~\ref{fig2}(left and center panels).
In the stripe phase, in addition to phase disorder, there is a disordering
of the orientational (``electron nematic'') order, resulting in a more visually
dramatic breakup into regions of vertical and horizontal stripes, as can be
seen in Fig.~\ref{fig1} (center panel).  

The effect of quenched disorder in the symmetric phase ($\alpha > 0$) is
somewhat different.  In a sense, the effect of the disorder is to pin the
fluctuating order of the proximate ordered phase.   However, here, whether
the disorder is weak or strong, it is nearly impossible to distinguish
fluctuating stripes from checkerboards.  Fig.~\ref{fig1} (right panel) and Fig.~\ref{fig2} (right panel) illustrate
this phenomenon.  This is easily understood in the weak disorder limit, where
\begin{equation}
\varphi_j(\rr) = \int d\rr^\prime \chi_0(\rr-\rr^\prime) e^{-i {\bf Q}\cdot \rr^\prime} U(\rr^\prime) + {\cal O}(U^3)
\label{weak}
\end{equation}
where the susceptibility,
\begin{equation}
\chi_0(\rr) = K_0(\sqrt{\alpha}\rr),
\end{equation}
is expressed in terms of the $K_0$ Bessel function and is independent of
$\gamma$.  Near criticality ($1\gg \alpha>0$), the susceptibility is very
long ranged, so a significant degree of local order can be pinned by even
a rather weak impurity potential.  However, only the higher order terms
contain any information at all about the sign of $\gamma$, and by the time
they are important, the disorder is probably already so strong that it blurs
the distinction between the two states, anyway.

\subsection{Diagnostic Filters}
\label{filters}
Now, our task is to answer the question:  Given a set of simulated data,
what quantitative criteria best allows us to infer the form of the relevant
correlations in the absence of disorder?  For sufficiently weak disorder,
these criteria are, at best, just a way of quantifying a conclusion that
is already apparent from a visual analysis of the data.  Where disorder is
of moderate strength, such criteria may permit us to reach conclusions that
are somewhat less prejudiced by our preconceived notions.  Of course, when
the disorder is sufficiently strong that the density-wave correlation length
is comparable to the CDW period, it is unlikely that any method of analysis
can yield a reliable answer to this question.

Firstly, to eliminate the rapid spatial oscillations, we define two scalar fields 
(which we will consider to be the two components of a vector field, $\bf{A}(\rr)$)
corresponding to the components of the density which oscillate, respectively,
with wave vectors near ${ Q \hat x}$ and $Q\hat y$:
\begin{equation}
{\rm A}_j(\rr) = \int \ud\rr^\prime F_j(\rr-\rr^\prime) \rho(\rr^\prime)
\end{equation}
where we take $F_j$ to be the coherent state with spatial extent equal to the CDW period:
\begin{equation}
F_j(\rr) = \frac{Q^2}{2\pi^2} \exp[i \,Q_j r_j - r^2/(2\pi\lambda^2)  ]
\end{equation}
(no summation over $j$.)

In terms of ${\bf A}$ we construct three quantities  which can be used
in interpreting data:  
\begin{eqnarray}
\xi_{CDW}^2 & \equiv & 
	\frac { \verts{\int \ud\rr  {\bf A} }^2} {\int \ud\rr \verts{\bf A}^2}  \label{xicdw}  \\
\xi_{orient}^2 & \equiv &  
	\frac{\verts{\int \ud\rr \left[ \verts{{\rm A}_1}^2 -  \verts{{\rm A}_2}^2 \right]}^2  } {\int \ud\rr \verts{ \verts{{\rm A}_1}^2  -  \verts{{\rm A}_2}^2  }^2 }  \label{xiorien} \\
\eta_{orient} & \equiv & 
\frac{\int \ud\rr  \verts{ \verts{{\rm A}_1}^2 - \verts{{\rm A}_2}^2 }^2 }
{ \int \ud\rr \verts{ \verts{{\rm A}_1}^2 + \verts{{\rm A}_2}^2 }^2  }   \label{horien}
\end{eqnarray}
The quantities called $\xi$ have units of length and $\eta_{orient}$ is
dimensionless.  All of these quantities are invariant under a change of
units, $\rho(\rr) \to \Lambda \rho(\rr)$;  this is important since in
many experiments, including STM, the absolute scale of the density
oscillations is difficult to determine because of the presence of unknown
matrix elements.

\begin{figure}
\includegraphics[height=0.62\textwidth,width=0.5\textwidth]{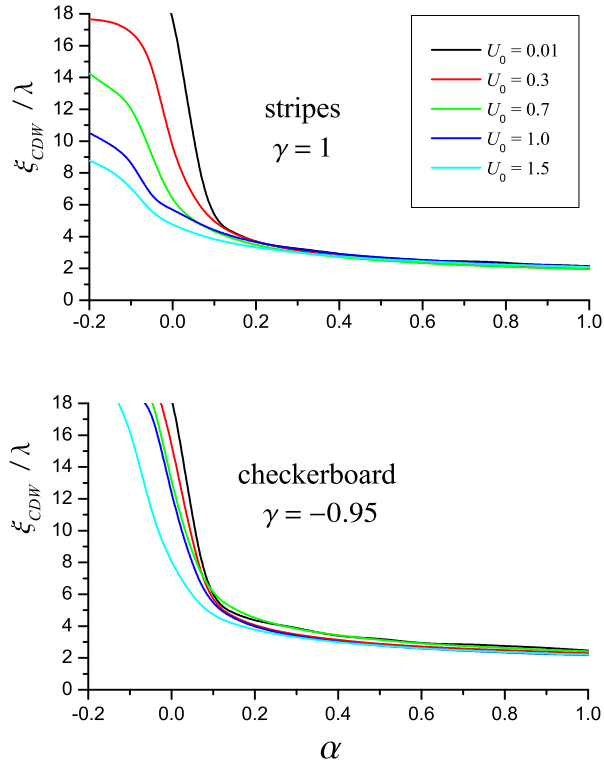}
\caption{\label{fig3}
(color online) $\xi_{CDW}$ vs. $\alpha$  (measured in units of the CDW wavelength).  
top panel) $\gamma = 1$,  bottom panel) $\gamma = -0.95$.  In a perfectly clean system,
$\xi_{CDW}$ vanishes for $\alpha > 0$, whereas with even a little disorder, charge
order is induced.  For $U_0 > 1$ and $\alpha<0$, disorder affects
$\xi_{CDW}$ more strongly in the stripe system.   For $\alpha>0$, there is little
distinction in either the sign of $\gamma$ or the strength of $U_0$.  All
quantities in Figs.~\ref{fig3},\ref{fig4} and \ref{fig5} are computed for systems
of size $20\lambda \times 20\lambda$ 
and averaged over 50 or more realizations of the disorder.}
\end{figure}

$\xi_{CDW}$ has the interpretation of a CDW correlation length.  In
the absence of quenched disorder, and for $\alpha <0$, $\xi_{CDW}\sim L$,
where $L$ is the linear dimension of the sample.  In the presence of
disorder, $\xi_{CDW}$ is an average measure of the domain size.
For $\alpha >0$ and weak but non-vanishing disorder, 
$\xi_{CDW}\sim \alpha^{-1/2}$, as can be seen from a scaling analysis
of Eq. \ref{weak}. 
The evolution of $\xi_{CDW}$ as a function of $\alpha$ is shown in Fig.~\ref{fig3}
for a system of size $L=20\lambda$, for various strengths of the disorder
and for stripes (Fig.~\ref{fig3} (top panel) and checkerboards (Fig.~\ref{fig3} (bottom panel).)
$\xi_{CDW}$ is generally a decreasing function of increasing
disorder, although for $\alpha >0$ there is a range in which it exhibits
the opposite behavior.  For fixed, non-zero disorder, we see that a
large value of $\xi_{CDW} > 4$ almost inevitably means that $\alpha <0$,
{\it i.e.} that the density patterns are related to a domain structure 
of what would otherwise have been a fully ordered state.  However, smaller
values of $\xi_{CDW}$ can either come from weak pinning of CDW order which
would otherwise  be in a fluctuating phase, or a very small domain structure
due to strong disorder.

\begin{figure}
\includegraphics[height=0.62\textwidth,width=0.5\textwidth]{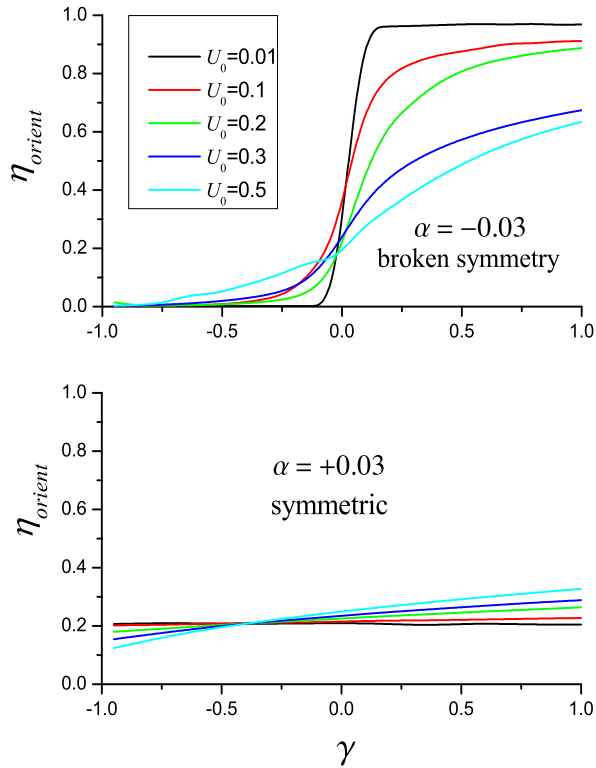}
\caption{\label{fig4}
(color online) $\eta_{orient}$ vs. $\gamma$: top panel)  $\alpha < 0$.  In the ordered phase, $\eta_{orient}$
is good indicator of the nature of the underlying order (i.e the sign of $\gamma$.
At large $U_0$, the distinction is lost, and the result approaches that of the
symmetric phase ($\alpha > 0$), shown in the bottom panel.  We observe that the nearly uniform
value of $\eta_{orient} \approx 0.2$ in the $\alpha > 0$ measurements 
intersects the (all of the) data in the ($\alpha = -0.03$) graph at the $\gamma = 0$ axis.
}
\end{figure}

The CDW correlation length does not distinguish between stripe and 
checkerboard patterns.  However, for $\alpha<0$, the orientational 
amplitude $\eta_{orient}$ is an effective measure of stripiness.  In 
the clean system with $\alpha<0$, $\eta_{orient}$ approaches unity for 
$\gamma > 0$ and is zero for $\gamma < 0$.  While quenched disorder
somewhat rounds the sharp transition in $\eta_{orient}$ at $\gamma=0$,
it is clear from Fig.~\ref{fig4} (top panel) that values of $\eta_{orient} > 0.2$ are clear 
indicators of stripe order, and $\eta_{orient} < 0.2$ implies checkerboard.  In the absence of disorder,
$\eta_{orient}$ is ill-defined for $\alpha >0$, and even for non-zero
disorder, the behavior of $\eta_{orient}$ is difficult to interpret in
the fluctuating order regime, as is also clear from Fig.~\ref{fig4}.  The
orientational correlation length, $\xi_{orient}$, gives similar
information as $\eta_{orient}$, and suffers from the same shortcomings.

\begin{figure}
\includegraphics[height=0.38\textwidth,width=0.49\textwidth]{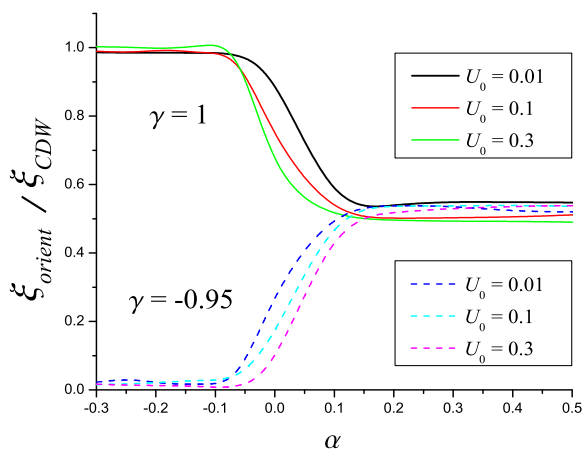}
\caption{\label{fig5}  
(color online) $\xi_{orient}/\xi_{CDW}$ vs. $\alpha$:  For $\alpha < 0$, $\xi_{orient}/\xi_{CDW}$
is a strong indicator of the sign of $\gamma$.  For $\alpha>0$ and either
sign of $\gamma$, the disorder-averaged ratio is $1/2$, largely independent
of other parameters.} 
\end{figure}

One interesting possibility is that, for a weakly disordered stripe
phase, one can imagine an orientational glass in which
$\xi_{orient} \gg \xi_{CDW}$, {\it i.e.} the CDW order is phase
disordered on relatively short distances, but the orientational
order is preserved to much longer distances.  In Fig.~\ref{fig5}, we plot
the ratio of $\xi_{orient}/\xi_{CDW}$ for $\gamma =1$ (strong preference
for stripes) as a function of $\alpha$ for various values of the
disorder.  Clearly, we have not found dramatic evidence of such an
orientational glass, although we have not carried out an exhaustive search.
Nonetheless, for $\alpha < 0$, this ratio is manifestly another good way
to distinguish stripe and checkerboard order.

The bottom line:  If $\xi_{CDW}$ is a few periods or more, it is
possible to conclude that $\alpha < 0$, {\i.e.} that in the absence
of impurities there would be long-range CDW order.  If $\xi_{CDW}$ 
is shorter than this, then either the impurity potential is very
strong (which should be detectable in other ways) or $\alpha\sim \xi_{CDW}^{-2}$
is positive.  For intermediate values of $\xi_{CDW}$,
all that can be inferred is that the system is near critical, $|\alpha| \ll 1$.
Given a substantial $\xi_{CDW}$, it is possible to distinguish a pinned 
stripe phase from a pinned checkerboard phase for which $\eta_{orient}$ is
greater than or less than 0.2, respectively.

\section{Effect of an orthorhombic distortion}
\label{ortho}

\begin{figure}
\includegraphics[height=0.31\textwidth,width=0.41\textwidth]{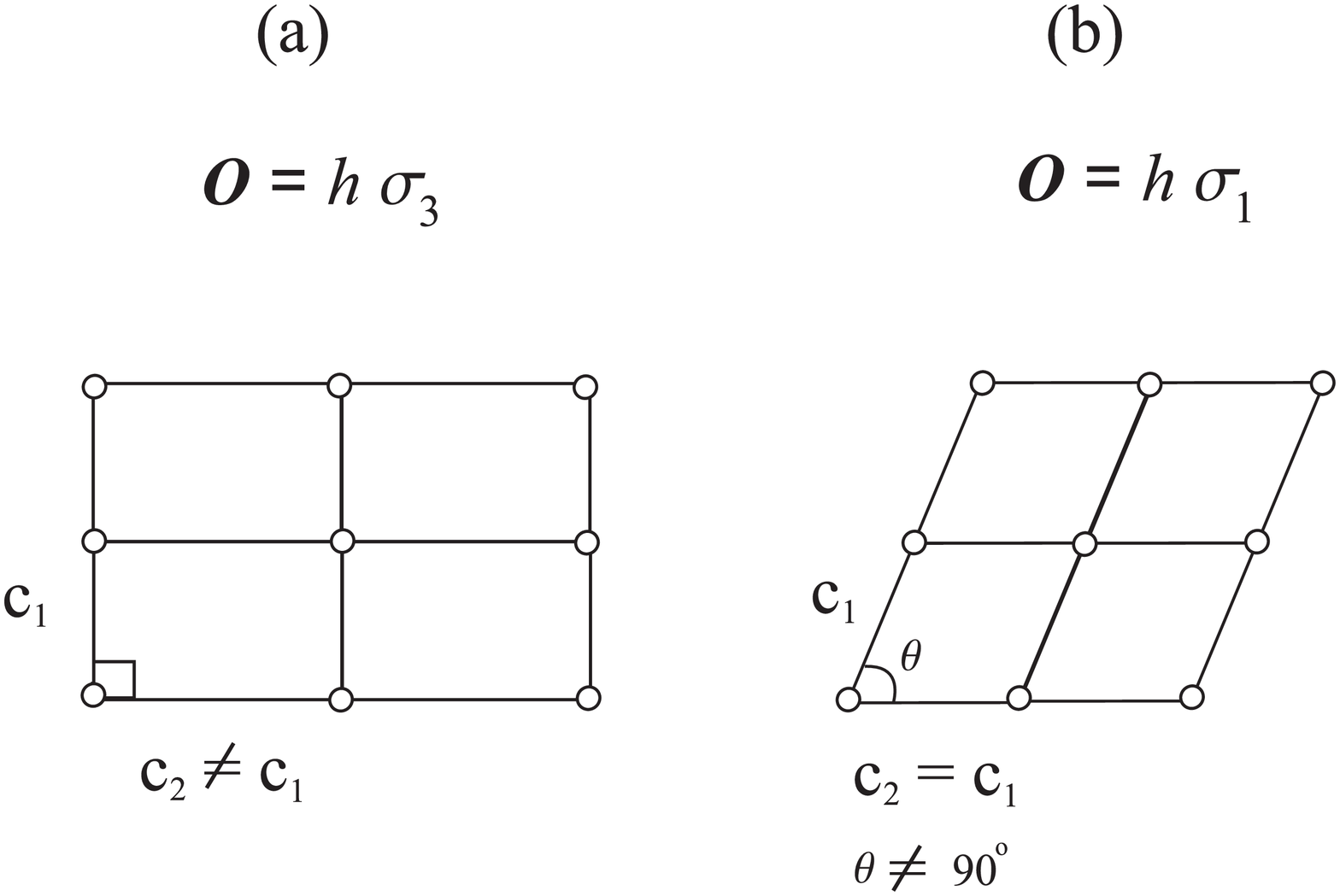}
\caption{\label{ortho_fig}
Orthorhombic symmetry breaking reduces a square lattice to a
lower symmetry.  (a)  Rectangular lattice distortion (exaggerated).  The preferred
orthorhombic axis lies along an original lattice vector (i.e. along the lines connecting
atomic sites.)  (b)  A rhombohedral distortion leaves the preferred orthorhombic
axis diagonal to the original lattice vectors.} 
\end{figure}

An orthorhombic distortion breaks the C$_4$ symmetry of the square lattice
down to C$_2$.  There are two distinct ways this can occur - either the
square lattice can be distorted to form rectangles, as shown in Fig.~\ref{ortho_fig}a,
in which case the ``preferred'' orthorhombic axis is either vertical or
horizontal, or the squares can be distorted to form rhombi, as shown in
Fig.~\ref{ortho_fig}b, in which case the preferred orthorhombic axis is diagonal.
A general orthorhombic distortion is represented by a traceless symmetric tensor,
$O_{ab}$;  an orthorhombic distortion corresponding to Fig.~\ref{ortho_fig}b is
represented by ${\bf O}=h{\bf \sigma}_3$ while Fig.~\ref{ortho_fig}b is 
${\bf O}=h{\bf \sigma}_1$ where $h$ is the magnitude of the symmetry breaking
and ${\bf \sigma}_j$ are the Pauli matrices.  Then
\begin{eqnarray}
H_{ortho} = -O_{ab}Q_aQ_b\left[|\vp_1|^2-|\vp_2|^2\right] \hspace{1.5cm} \nonumber \\
+ \ \ g\left[Q_aO_{ab}\vp_1^\star\partial_b\vp_1 -\epsilon_{a\bar a}Q_aO_{\bar a b}\vp_2^\star\partial_b\vp_2\right] +\ldots
\end{eqnarray}
where $\ldots$ is higher order terms.

In case (a), the first term is non-zero, and hence dominant.  For $h$ positive,
this enhances $\vp_1$ and suppresses $\vp_2$.  In a stripe phase, this has
the same effect as a magnetic field in a ferromagnet - it chooses among the
otherwise degenerate vertical and horizontal stripe ordered states, so one is
preferred.\cite{carlson1}  For checkerboard order, it produces a
distortion of the fully ordered state, so that the expectation value of
$\vp_1$ exceeds the expectation value of $\vp_2$.  Moreover, it results
in a split phase transition, so that as a function of decreasing temperature,
rather than a single transition from a symmetric high temperature phase to
a low temperature checkerboard phase, in the orthorhombic case there are two
transitions, the first to a stripe ordered phase, and then at a temperature
smaller by an amount proportional to $h$, a transition to a distorted
checkerboard phase.  The second term, proportional to $g$, is subdominant
in this case, but still has a significant effect.  For an incommensurate
stripe phase, it results in a small shift in the ordering wave vector 
${\bf Q}\to \tilde {\bf Q}={\bf Q}(1-gh/\kappa_L)$.  In an incommensurate
checkerboard phase, it results in a relative shift of the two ordering vectors,
${\bf Q}\to \tilde {\bf Q}={\bf Q}(1-gh/\kappa_L)$ and
${\bf Q}^\prime\to \tilde {\bf Q}^\prime={\bf Q}^\prime(1+gh/\kappa_L)$ one toward
smaller and the other toward larger magnitude producing a rectangular checkerboard.
In the case in which the order is commensurate, it is locked to the lattice, and
therefore the only shifts in ordering wave-vectors are proportional to the
(usually miniscule) shifts of the lattice constant.  

In case (b), the first term vanishes, so the second term is dominant.  For
incommensurate order, this results in a small {\it rotation} of the ordering
vector away from the crystalline symmetry axis.  To first order in $h$, the
new ordering vector is $\tilde {\bf Q} =|{\bf Q}| \langle 1,k\rangle$ with
$k=gh/\kappa_T$ and, in the case of checkerboard order, the second ordering
vector is $\tilde {\bf Q}^\prime=|{\bf Q}| \langle k,1\rangle$.  Again, in
the commensurate case, the order remains locked to the lattice until the
magnitude of the orthorhombicity exceeds a finite critical magnitude.

To summarize, the response of charge order to small amounts of orthorhombicity
can be qualitatively different depending on whether the order is commensurate
or incommensurate and checkerboard or striped. 

\subsubsection{More complex patterns of symmetry breaking}

It is useful to point out that with complex crystal structures, the
application of the above ideas requires some care.  For example, there
are some cuprate superconductors which exhibit a so called Low Temperature
Tetragonal (LTT) phase.  This phase has an effective orthorhombic
distortion of each copper oxide plane, but has two planes per unit cell
and a four-fold twist axis which is responsible for the fact that it is
classified as tetragonal.  In the first plane, ${\bf O}= h{\bf \sigma}_3$,
and in the second ${\bf O}=- h{\bf \sigma}_3$.  Note that this means that
for stripe order, there will be four ordering vectors, a pair at 
$\tilde {\bf Q}=\pm |{\bf Q}|\langle 1+hg/\kappa_L,0\rangle$ from the first
plane and a pair at $\tilde {\bf Q}=\pm |{\bf Q}|\langle 0, 1+hg/\kappa_L\rangle$
from the other.  However, for incommensurate checkerboard order, there should
be {\it eight} ordering vectors: 
$\pm |{\bf Q}|\langle 1+hg/\kappa_L,0\rangle$, $\pm |{\bf Q}|\langle 1-hg/\kappa_L,0\rangle$, $\pm |{\bf Q}|\langle 0, 1+hg/\kappa_L \rangle$, and $\pm |{\bf Q}|\langle 0, 1-hg/\kappa_L\rangle$.

\section{Analysis of experiments in the cuprates}
\label{exp}

There have been an extremely large number of experiments which have been performed on
various closely related cuprates, both superconducting and not, which have been
interpreted as evidence for or against the presence of charge order of various
types.  For instance, there is a large amount of quasi-periodic structure observed
in the local density of states measured by scanning tunneling microscopy (STM) on
the surface of superconducting Bi$_2$Sr$_2$CaCu$_2$O$_{8+\delta}$ crystals, but
there is controversy concerning how much of this structure arises from the 
interference patterns of well-defined quasiparticles whose dispersion is determined
by the d-wave structure of the superconducting gap~\cite{hoffman1,mcelroy1,wang1,bena1} 
and how much reflects the presence of charge order or incipient charge 
order.\cite{howald1,hoffman2,howald2,mcelroy2,momono1,kivelson1}  A similar debate has been carried out
concerning the interpretation of the structures seen in inelastic neutron-scattering
experiments.\cite{tranquada1,tranquada2,tranquada3,li1,keimer1,bourges1,mook1,christensen1,kivelson1}    

As mentioned in the introduction, the issue of how to distinguish charge order
from interference patterns was discussed in detail in a recent review,\cite{kivelson1}
and so will not be analyzed here.  Here, we will accept as a working hypothesis
the notion that various observed structures should be interpreted in terms of
actual or incipient order, and focus on identifying the type of order involved.

\subsection{Neutron and X-ray scattering}
\label{scattering}

Scattering experiments in several of the cuprates, most notably {\LSCO},
{\LNSCO}, {\LBCO}, and O-doped {\LCO} have produced clear
and unambiguous evidence of charge and spin ordering phenomena with a characteristic
ordering vector which changes with 
doping.\cite{yamada1,cheong1,tranquada4,thurston1,mason1,yamada2,tranquada5,tranquada6,stock1,stock2}
The evidence is new peaks in the static
structure factor corresponding to a spontaneous breaking of translational
symmetry, leading to a new periodicity longer than the lattice constant of
the host crystal.  In many cases, the period is near 4 lattice constants
for the charge modulations and 8 lattice constants for the spin.  
The peak-widths correspond to a correlation length~\cite{lee1,tranquada6} that is
often in excess of 20 periods.  For technical reasons, the spin-peaks are easier to detect
experimentally, but where both are seen, the charge ordering peaks are always
seen~\cite{wochner1,abbamonte1,zimmermann1,fujita2,reznik1} to be aligned with the spin-ordering peaks, and the charge period
is 1/2 the spin period.

Except in the case~\cite{wakimoto1,fujita3} of a very lightly doped ($x < 0.05$)
LSCO (where the stripes lie along an orthorhombic symmetry axis, so only
two peaks are seen), there are four equivalent spin-ordering peaks and, where
they have been detected,
four equivalent charge ordering peaks.  Thus, the issue arises whether this
should be interpreted as the four peaks arising from some form of checkerboard
order, or as two pairs of peaks arising from distinct domains of stripes - half
the domains with the stripes oriented in the $x$ direction and half where they
are oriented along the $y$ direction.  A second issue that arises is whether
the charge order is locked in to the commensurate period, 4, or whether it is
incommensurate.

A variety of arguments that the scattering pattern is revealing stripe
order, and not checkerboard order, were presented in the original paper by
Tranquada {\it et.~al.}\cite{tranquada-private} (and additionally in Ref.~\onlinecite{zimmermann1,tranquada7}) 
where the existence of charge order in a cuprate high temperature superconductor was first
identified.  Here, we list a few additional arguments based on the symmetry analysis
performed in the present paper, which support this initial identification:
\noindent{\bf 1.}  It follows from simple Landau theory~\cite{zachar1} that
if there is non-spiral spin-order at wave-vectors $\vec Q_i,\vec Q_j$, there will necessarily
be charge order at wave-vectors $\vec Q_i + \vec Q_j$.  Thus, if the four
spin-ordering peaks come from checkerboard order, then charge-ordering peaks
should be seen at wave vectors $\pm 2\vec Q_1$, $\pm 2\vec Q_2$ and
$\pm\vec Q_1\pm \vec Q_2$, while if they come from stripe domains of the two
orientations, no peaks at $\pm\vec Q_1\pm \vec Q_2$ should be seen.  The latter
situation applies to all cases in which charge ordering peaks have been seen at all.
\noindent{\bf 2.}  As mentioned above, in the LTT phase, the crystal fields
should cause small splittings of the ordering vectors in an incommensurate
checkerboard phase, causing there to be eight essentially equivalent Bragg
peaks, as opposed to the four expected for domains of stripes of the two
orientations.  No such splittings have been detected in any of the scattering
experiments on {\LNSCO} and {\LBCO} crystals consistent with stripe domains.
\noindent{\bf 3.}  It should be mentioned that the fact that the LTT phase
stabilizes the charge order is, by itself, a strong piece of evidence that
the underlying charge order is striped.  In this phase, the O octahedra are
tipped in orthogonal directions in alternating planes, and the direction of
the tip is along the Cu-O bond direction.  This permits a uniquely strong
coupling between the octahedral rotation and  stripe order.\cite{abbamonte1,fujita1,fujita2,kampf1}  

A second issue, especially when the period of the charge order is near
4 lattice constants, is whether the charge order is commensurate or
incommensurate.  One way to determine this is from the position of the
Bragg peak - in the commensurate case, the structure factor should be
peaked at $2\pi/4a$ ($2\pi/8a$ for the spin order), and should be locked
there, independent of temperature, pressure, or even doping for a finite
range of doping.  Most of the reported peaks seen in scattering are not
quite equal to the commensurate value, however.  
In the LTT phase of {\LBCO}, it is believed the stripe phase is locally
commensurate.  The ordering wave vector is temperature independent in the
LTT phase, but jumps at the LTT-LTO transition and continues to change on
warming.  For LSCO in the LTO phase, the stripes might be incommensurate,   
however, there are only 4 peaks seen and not 8.  So it must be incommensurate
stripe order and not checkerboard order.\cite{comment-referee}
A clearer piece of evidence comes from the rotation of the ordering
vector away from the Cu-O bond direction in the LTO phase of {\LSCO} and
O doped {\LCO}.  In both cases, there is a small angle rotation (less than 4\degree)
seen, which moreover decreases with doping as the magnitude of the
orthorhombic distortion decreases.\cite{fujita1}  As discussed above, this
is the generic behavior expected of incommensurate order, and is incompatible
with commensurate order.

\subsection{STM}
\label{stm}

The strongest quasiperiodic modulations seen in STM are those reported
by Hanaguri {\it et.~al.}~\cite{hanaguri1} on the surface of NaCCOC, which
have a period which appears to be commensurate, $4\,a$.  This observation
has been interpreted as evidence that NaCCOC is charge-ordered
with a checkerboard pattern (at least at the surface.\cite{brown1}) 
However, the correlation length deduced for the checkerboard order is only
about two periods of the order.  Indeed, the domain structure in the STM
data looks to the eye very much like the
pictures in our Figs.~\ref{fig1} (right panel) and~\ref{fig2} (right panel).
 This suggests the
possibility that: {\bf  1:} What is being seen is pinning of what, in the
disorder free system, would be fluctuating order ($\alpha > 0$) relatively
close to the quantum critical point.
{\bf 2:}  That the nearby ordered state could be either a striped or a
checkerboard state.
We hope, in the near future, to apply the more quantitative analysis
proposed in the present paper to this data.

\begin{figure}
\includegraphics[height=0.4\textwidth,width=0.47\textwidth]{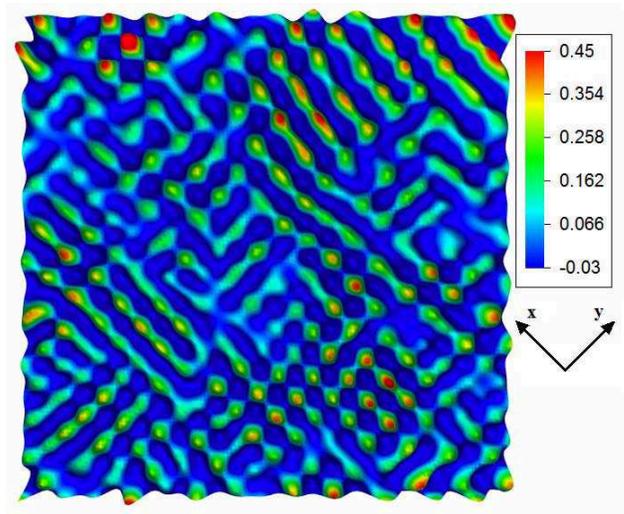}
\caption{\label{alan} 
(color online) LDOS integrated in energy up to $E=+15meV$. [Color scale is arbitrary.]
Both $\xi_{orient}$ and $\xi_{CDW}$ are quite small, suggestive that
the system is in a disorder-pinned, fluctuating phase.} 
\end{figure}

\emph{Concerning the modulations seen in STM studies on BSCCO}:
Given the recent interest in Bi$_2$Sr$_2$CaCu$_2$O$_{8+\delta}$, we report
a preliminary application of our analysis to data from a near optimally
doped sample, with an
image size 21 CDW wavelengths across.  Fig.~\ref{alan} is a map of the LDOS
integrated
in energy to +15meV.~\cite{data_note}  (The axes here are rotated
45\degree~relative to those in
Figs.~\ref{fig1} and~\ref{fig2}.)
In producing Fig.~\ref{alan}, we employ a Fourier mask (such as the one used
Ref.~\onlinecite{fang1}) as
a visual aid to show that there are indeed period 4 oscillations.  This is a
coherent state filter, centered
in Fourier space around $2\pi/a (\pm 1/4,0)$ and $2\pi/a (0,\pm 1/4)$, and
with a wide, flat top.
Using the Eqns.~\ref{xicdw}-\ref{horien}, we find
$\xi_{orient} = 4.5 \lambda$ and $\xi_{CDW} = 2.5 \lambda$, with $\lambda
\approx 4.2 a$,
and $\eta_{orient} = 0.28$, which corresponds to $\gamma \gtrapprox 1/2$ and
relatively strong
disorder ($U_0 \approx 0.5$).
Additional measurements of the (unintegrated) LDOS on the
same sample at $E=8meV,15meV$ yield comparable correlation lengths. From
these we conclude
the system shows a short-ranged mixture of (disorder-pinned) stripe and
checkerboard order,
and in the absence of pinning, would be in its fluctuating (symmetric)
phase, but close to
the critical point ($\alpha$ small). (Though there should probably be a fair
amount of
quasiparticle scattering at a nearby wave vector, it should be four-fold
symmetric,
so should not affect either $\xi_{orient}$ or $\eta_{orient}$.)
The fact that the orientational correlation length exceeds the CDW
correlation length
is suggestive that the proximate ordered state is a stripe ordered state and
the
ratio $\xi_{orient}/\xi_{CDW} \approx 2$ is interesting, as it exceeds our
(disorder-averaged) result of 1/2 for the symmetric phase ($\alpha > 0$).
However, undue weight should not be given to this result,
as the ($\alpha > 0$) region of Fig.~\ref{fig5} is a product of
disorder-averaging,
and Fig.~\ref{alan} is a single set of data.  In the future, we hope to
apply our
methods to a more substantial set of experimental data.


\section{Conclusions}
\label{conclude}

There are
many circumstances in which charge order plays a significant role in the physics
of electronically interesting materials.  Depending on the situation, different
aspects of the physics may be responsible for the choice of the characteristic
period of the charge order;  for instance, it can be determined by Fermi surface
nesting (as in a Peierls transition), by a small deviation from a commensurate
electron density (which fixes a concentration of discommensurations), or by some
form of Coulomb frustrated phase separation.  Working backwards, measurements of
the period of the charge order as a function of parameters (temperature, pressure,
doping, \ldots) can shed light on the mechanism of charge ordering.

The physics that determines the ultimate pattern of charge order is still more
subtle.  For instance, for adsorbates on graphite, the sign of the energy of
intersection determines whether the discommensurations form a striped or honeycomb
arrangement.~\cite{coppersmith1}  In 2H-TaSe$_2$, broken hexagonal symmetry has been 
observed~\cite{fleming1} in x-ray scattering and TEM~\cite{onozuka1} (such
a system has been studied by McMillian~\cite{mcmillan1} using LG methods.) 
In certain nearly tetragonal rare-earth tellurides, which have
been found to form stripe ordered phases,~\cite{laverock1,dimasi1} this can be shown to be a
consequence of some fairly general features of the geometry of the nested portions
of the Fermi surface so long as the transition temperature is sufficiently high.\cite{yao1}  

In the cuprates, calculations of the structures originating from Coulomb-frustrated
phase separation,\cite{low1} DMRG calculations on t-J ladders,\cite{white1} and Hartree-Fock
calculations on the Hubbard model\cite{zaanen1,schulz1,machida1} all suggest
that stripe order is typically preferred over checkerboard order.  
Conversely, the Coulomb repulsion between dilute doped holes,
or between dilute Cooper pairs favor a more isotropic (Wigner crystalline)
arrangement of charges with more of a checkerboard structure.\cite{chen1,komiya1,fu1,fine04}  Thus, resolving
the nature of the preferred structure of the charge ordered states in the cuprates,
at the least, teaches us something about the mechanism of charge ordering in these
materials.  

On the basis of our present analysis, we feel that there is compelling evidence
that most, and possibly all, of the charge order and incipient charge order
seen in hole-doped cuprates is preferentially striped.  We also conclude that most
of the structure seen in STM studies is disorder pinned versions of what would, in
the clean limit, be fluctuating stripes, rather than true, static stripe order.

Note: After this work was completed we received a draft of a paper by del Maestro and coworkers\cite{delmaestro1} who discuss similar ideas to the ones we present in this paper.  We thank these authors for sharing their work with us prior to publication. After this paper was submitted for publication M. Vojta pointed out to us that in a very recent paper he and his coworkers considered the effects of slow thermal fluctuations of stripe and checkerboard charge order on the magnetic susceptibility of disorder-free high $T_c$ cuprates.\cite{Vojta05}.

Note added: While this paper was being refereed a new neutron scattering study of LNSCO became available\cite{christensen2}, which confirmed the existence of unidirectional charge order (stripe) and collinear spin order in this material, in agreement with the results and interpretation of Ref.[\onlinecite{tranquada4}].

\begin{acknowledgments}
The authors would like to thank P. Abbamonte, J. C. Davis, R. Jamei, E-A. Kim, S. Sachdev, 
J. Tranquada for many useful discussions.  This work was supported through the
National Science Foundation through grant Nos. NSF DMR 0442537 (E.~F., at the University of Illinois)
and NSF DMR 0531196 (S.~K. and J.~R., at Stanford University) and through the Department
of Energy's Office of Science through grant Nos. DE-FG02-03ER46049
(S.~K., at UCLA), DE-FG03-01ER45925 (A.~F. and A.~K., at Stanford University), and DEFG02-91ER45439, through the Frederick Seitz Materials Research Laboratory at the University of 
Illinois at Urbana-Champaign (EF).
\end{acknowledgments}


\begin{thebibliography}{69}
\bibitem[{com()}]{comment-referee}
\bibinfo{note}{This statement can be made more concrete by the following line
  of reasoning. The splitting of the peaks is required by symmetry, but the
  magnitude of the splitting is determined by microscopic considerations. There
  is certainly a locally orthorhombic pattern of O octahedral tilts. These, in
  turn, certainly affect the magnitude of the hopping matrix elements, making
  the hopping matrix larger in one direction and smaller in the other. In turn,
  this causes a distortion in the band structure. Were we to imagine that the
  charge order was associated with Fermi surface nesting, then the difference
  between the ordering vectors in the two directions would be proportional to
  the distortion of the hopping matrix elements, which could be big even for a
  rather modest lattice distortion. However, if the charge order were
  commensurate with the lattice, i.e. if the order is locked to the lattice
  periodicity, then only a direct distortion of the lattice periodicity in the
  two orthogonal directions could split the Bragg peak, and hence this
  splitting would be unobservably small. Thus the 4 vs 8 peak argument probably
  cannot be used to distinguish stripes from checkerboards in the case of
  commensurate order, but with high probability (depending, of course, on
  microscopic details) can be used to distinguish them when the order is
  incommensurate.}
\bibitem[{\citenamefont{{This data was taken in conjuction with Craig Howald at
  Stanford University.}}()}]{data_note}
\bibinfo{author}{\bibnamefont{{This data was taken in conjuction with Craig
  Howald at Stanford University.}}}
\end{thebibliography}
\end{document}